\documentclass[a4paper,11pt]{article}
\usepackage{jcappub} 
\usepackage{lineno}
\newcommand{\logg}{\log_{10}\left[G_\text{eff}\text{MeV}^2\right]}

\title{\boldmath First constraints on non-minimally coupled Natural and Coleman-Weinberg inflation and massive neutrino self-interactions with Planck+BICEP/Keck}

\author[a]{Nilay Bostan}
\author[b]{and Shouvik Roy Choudhury}

\affiliation[a]{Proton Accelerator Facility, Turkish Energy Nuclear and Mineral Research Agency, Nuclear Energy Research Institute, 06980, Ankara, Türkiye}
\affiliation[b]{Institute of Astronomy and Astrophysics, Academia Sinica, No. 1, Section 4, Roosevelt Road, Taipei 106216, Taiwan}

\emailAdd{nilay.bostan@tenmak.gov.tr}
\emailAdd{sroy@asiaa.sinica.edu.tw}

\abstract{In this work, for the first time in literature, we study the predictions of non-minimally coupled Natural and Coleman-Weinberg potentials in the $n_s-r$ plane, and an extended $\Lambda$CDM model where we include non-standard self-interactions among massive neutrinos, mediated by a heavy scalar or vector boson. Constraints were derived using the Planck 2018 + BICEP/Keck 2018 datasets along with other data. For the inflationary potentials, we consider two different formulations in gravity that are non-minimally coupled to the scalar field of the inflaton: \textit{Metric and Palatini.} We only consider the self-interaction to be present among $\tau$-neutrinos and only at moderate strengths. This is because strong interactions among $\tau$-neutrinos, or any strength self-interaction among electron- and muon-neutrinos, as well as any strength flavor-universal interactions, are strongly disfavoured from particle physics experiments. 

In terms of cosmological data, we use the latest public CMB datasets from Planck 2018 and BICEP/Keck 2018 collaborations, along with other data from CMB lensing, BAO, RSD, and SNe Ia luminosity distance measurements. We find that there are some situations where predictions from the inflationary models are ruled out at more than 2$\sigma$ by the minimal $\Lambda$CDM$+r$ model, but they are allowed in the self-interacting neutrino scenario.}

\begin{document}

\maketitle
\flushbottom

\section{Introduction}
\label{sec:intro}
A well-established model for the early universe is the inflationary scenario that explains very well the horizon and flatness problems, as well as provides very precise predictions of the primordial density fluctuations, which can be verified by cosmological observations~\cite{Guth:1980zm,linde1982new,albrecht1982cosmology,linde1983chaotic}. A slow-rolling scalar field ($\phi$) known as the inflaton with a flat potential $V(\phi)$, ensures a straightforward mechanism for how inflation can happen. In literature, the majority of the inflationary models that have been examined up to now, are based on inflaton~\cite{martin2014encyclopaedia}. In addition, the inflationary parameters, $n_s, r$, can be computed and compared with constraints, which are obtained from measurements of the cosmic microwave background (CMB) anisotropies~\cite{akrami2020planck,BICEP:2021xfz,aghanim2020planck}. For instance, the observational parameters, such as the scalar spectral index, $n_s$ is constrained by Planck data to $0.965\pm0.004$~\cite{aghanim2020planck} in the $68\%$ confidence level region in the $\Lambda$CDM model, and the amplitude of tensor perturbations that is the tensor-to-scalar ratio, $r$ has just recently constrained by BICEP/Keck data to $r < 0.036$~\cite{BICEP:2021xfz} in the 95\% CL. 

Furthermore, there have been many proposed inflationary models so far, but a large number of them have already been ruled out by the cosmological observations, such as $\phi^2$ and $\phi^4$ potential models in which the inflaton is minimally coupled~\cite{akrami2020planck}. In fact, in curved space-time, the renormalizable scalar field theory necessitates the non-minimal coupling term, which is $\xi \phi^2 R$, between the inflaton and Ricci scalar~\cite{callan1970new, freedman1974energy,buchbinder1992effective}. This term makes most inflationary models more compatible with the observations, such as the Starobinsky ($R^2$ inflation) model~\cite{kehagias2014remarks}, which gives the best consistency for all current existing data. Furthermore, it is pivotal to mention that refs.~\cite{Buchbinder:1985jxp,Elizalde:1994im,Elizalde:1993qh,Elizalde:1993ew,Elizalde:1993ee,Elizalde:1994gv} considered and calculated with details the quantum corrections to scalar potential with the existence of non-minimal scalar-curvature term. Moreover, in literature, inflation with a non-minimally coupled scalar field has been studied for two different formulations in gravity~\cite{bezrukov2011higgs,bezrukov2008standard,Bostan:2018evz,bauer2008inflation,tenkanen2017resurrecting,rasanen2017higgs,jinno2020higgs,rubio2019preheating,enckell2018higgs,bostan2020quadratic,tenkanen2019minimal,jarv2020equivalence,Bostan:2019fvk,Bostan:2022swq,Bostan:2019uvv,Racioppi:2019jsp,Gialamas:2023flv,Eadkhong:2023ozb, Racioppi:2017spw,Okada:2022yvq,Choudhury:2017cos,Choudhury:2013zna,Choudhury:2014sxa,Choudhury:2015baa,Boubekeur:2015xza,Cecchini:2024xoq}, \textit{Metric and Palatini formulations.} In the Einstein-Hilbert Lagrangian, equations of motion are the same for both formulations. This means that two formalisms correspond to the same physics theories. On the one hand, these two formulations explain two different theories of gravity on the condition that there is a non-minimal coupling between matter and gravity~\cite{bauer2008inflation}. In this case, these two formulations are not equivalent at all. Also, the inflationary predictions of the two formulations are different for the potential models that are taken into account. For instance, in particular, the Starobinsky model is lost in the Palatini formulation, and in the Palatini formulation, $r$ can take much smaller values than the Metric one for large values of the coupling parameter between the inflaton and gravity ~\cite{bauer2008inflation, jarv2018palatini}. However, in the minimal coupling scenario, both metric and Palatini formulations are basically same and yield similar predictions of inflationary parameters. 

In addition, neutrinos play an important role in the evolution of the universe. Neutrinos contribute to the radiation energy density in the early universe, but in the late universe when they become non-relativistic (as the temperature falls below their masses) they contribute to the matter budget. Neutrino mass (total mass, $\sum m_\nu$) and energy density (parameterized with $N_{\rm eff}$, the effective number of relativistic degrees of freedom other than photons) are known to affect the inflationary parameters like the scalar spectral index, $n_s$ (while bounds on the $r$ are usually almost model-independent). Compared to the vanilla $\Lambda$CDM+$r$ model, both neutrino parameters are known to expand the allowed $n_s-r$ parameter space, with the dominant effect coming from $N_{\rm eff}$ \cite{Gerbino:2016sgw}. The effects of possible non-standard interactions in the neutrino sector on the inflationary parameters were previously studied in \cite{choudhury2022massive} with the latest Planck 2018 likelihoods \cite{Planck:2018vyg} and CMB B-mode BICEP/Keck likelihoods \cite{BICEP:2021xfz}. See also \cite{Barenboim:2019tux} for a study with the older Planck 2015 likelihoods \cite{Planck:2015fie}. These previous works looked at the viability of Natural inflation and Coleman-Weinberg inflation in the minimal coupling scenario ($\xi = 0$) in the presence of non-standard neutrino self-interactions mediated by a heavy scalar (with mass $> 1$ keV). For the modes probed by the CMB, the scalar particle can be considered to have decayed away, and the interaction can be approximated by a 4-fermion interaction with an effective coupling strength $G_{\rm eff}$\footnote{See~\cite{RoyChoudhury:2020dmd,choudhury2022massive,Choudhury:2021dsc,Oldengott:2017fhy,Barenboim:2019tux,Archidiacono:2013dua,Cyr-Racine:2013jua,Lancaster:2017ksf,Kreisch:2019yzn,Brinckmann:2020bcn,Das:2020xke,Mazumdar:2020ibx,Das:2023npl,Camarena:2023cku,He:2023oke,Lu:2023uhc} for previous studies on cosmological constraints on $G_{\rm eff}$. There are however other scenarios, i.e., when $m_\Phi \sim T$ or smaller there is a significant change in the phenomenology of the model: The system undergoes recoupling instead of decoupling, and thus the effect of the $\Phi$ particles in the background evolution cannot be ignored. We refer the reader to e.g.\ Refs.\ \cite{Archidiacono:2013dua,Forastieri:2015paa,Song:2018zyl,Forastieri:2019cuf,Archidiacono:2020yey,Escudero:2021rfi,Escudero:2019gvw,Corona:2021qxl,Venzor:2022hql,Sandner:2023ptm} for a more detailed discussion. See also \cite{RoyChoudhury:2018bsd,Chacko:2020hmh,Taule:2022jrz,Blinov:2020hmc,Escudero:2019gfk,Park:2019ibn,Chang:2022aas,Das:2022xsz,Esteban:2021tub,Huang:2021dba,Sung:2021swd,Mazumdar:2019tbm,Esteban:2021ozz,Kharlanov:2020cti,Venzor:2020ova,Venzor:2023aka,Dhuria:2023yrw,Dhuria:2023itq,Fiorillo:2023ytr,Fiorillo:2023cas,Fiorillo:2022cdq,Akita:2021hqn} for discussions in the related fields.}. One of the main effects of such an interaction on the inflationary parameters is the preference for lower $n_s$ values which brings the predictions from the minimal coupling versions of Natural and Coleman-Weinberg inflation models within 2$\sigma$ of the allowed $n_s-r$ region, even though these models are ruled out at more than 2$\sigma$ in the vanilla $\Lambda$CDM$+r$ model.

Inflationary models assure a highly motivated explanation for the large scale isotropy, homogeneity, and flatness of our universe, as well as the CMB anisotropies and large-scale structure of our universe ~\cite{Guth:1980zm,linde1982new,albrecht1982cosmology,linde1983chaotic, martin2014encyclopaedia}. In particular, the origin of the universe, the inflation dynamics and evolution, and the origin of primordial density perturbations can be understood very clearly with the Coleman-Weinberg mechanism  ~\cite{linde1982new,albrecht1982cosmology, coleman1973radiative, albrecht1985realization, linde1990particle, Shafi:1983bd, Lazarides:1984pq}. In addition, the combination of CW mechanism with the BSM physics can help differentiate between various models of inflationary potentials ~\cite{martin2014encyclopaedia}. Also, the Coleman-Weinberg mechanism is a quantum correction to the effective potential in the QFT framework ~\cite{coleman1973radiative}. It is essential to comprehend how mass is created and how spontaneous symmetry breaking occurs in the elementary particles~\cite{Weinberg:1973am, freedman1974energy}. Beyond the Standard Model's Higgs mechanism, BSM theories mostly propose new scalar fields and symmetry-breaking mechanisms. A framework to analyze and understand the dynamics of these new fields and their interactions is provided by the Coleman-Weinberg mechanism, which may lead to new phenomena and new windows in particle physics ~\cite{Shafi:1983bd, Lazarides:1984pq}. On the other hand, Natural inflation is an important inflationary potential model, which causes an inflationary period in the early universe \cite{martin2014encyclopaedia}. Also, it has a very simple form and doesn't have the \textit{eta problem} that corresponds to fine-tuning problems pertaining to the steepness of the inflationary potential, and does not affect Natural inflation, if compared with the other inflationary potential models \cite{martin2014encyclopaedia, freese1990natural, adams1993natural}. In this inflationary model, the inflaton field causes the potential to be naturally flat, negating the requirement for parameter fine-tuning. Thus, investigating Natural inflation can provide important information about the early universe. In particular, the important models in particle physics, especially the models involving axion-like particles, can ensure good motivation for Natural inflation itself. In addition to this, in Natural inflation the inflaton field naturally arises as a pseudo-Nambu-Goldstone boson and the field is the axion-like inflaton ~\cite{freese1990natural}. Furthermore, axion-like particles could be part of the scalar field that drives inflation in the early universe. Importantly, axions are hypothetical particles that appear in theories to solve QCD strong CP problems~\cite{Peccei:2006as}, as well as axions are the serious candidates of dark matter with strong theoretical motivations~\cite{Adams:2022pbo}. It can be mentioned that the particle physics models naturally occur in supersymmetry, unification, string theory, and cosmology, which makes Natural inflation an attractive way to investigate the relationship between them. As a result, Natural inflation provides a useful theoretical framework that combines cosmology and elementary particles in physics within the BSM context, and this provides a very important perspective in terms of building a bridge to physics beyond the standard model~\cite{Furuuchi:2013ila, Yonekura:2014oja}.

While the metric formulation of gravity corresponds to the standard general relativity, the motivation to study the Palatini formulation of gravity comes from the fact that it creates new opportunities for the solutions of modified gravity models and their cosmological implications~\cite{bauer2008inflation}. For instance, to understand the origin, evolution and nature of dark matter, dark energy, etc, BSM theories are very good alternatives to probe changes in gravity models. Now, while the minimal coupling scenarios of Natural and CW inflation are ruled out at more than 2$\sigma$ in the $\Lambda$CDM model, with the presence of non-minimal coupling to gravity can have important consequences on the predicted inflationary parameters depending on the coupling strength~\cite{Bostan:2020pnb, Bostan:2022swq, reyimuaji2021natural, Shafi:2006cs, Bostan:2018evz}. The Palatini formulation of gravity ensures a noteworthy path to examine these alterations and how they affect cosmological models (especially the inflationary observable parameters) in the case when there is a presence of non-minimal coupling between the inflaton and gravity~\cite{bauer2008inflation, Bostan:2019uvv, Bauer:2010jg}. And thus, it is important to take into account both metric and Palatini formulations of gravity, especially while investigating the cosmological parameters because these parameters have different values for these two formulations of gravity~\cite{bauer2008inflation, Bauer:2010jg}. 

As for neutrinos which are the most abundant massive particles in the universe, if they have self-interactions, it might have significant outcomes from the BSM perspectives as well~\cite{Berryman:2022hds, Choudhury:2021dsc}. Because of their mass squared differences, they oscillate during their travel through space. The nature and dynamics of these neutrino oscillations, neutrino propagation in high energy astrophysical phenomena, and large-scale structure of the universe could all be impacted by self-interacting neutrinos themselves~\cite{Berryman:2022hds}. Furthermore, the role of neutrinos is also pivotal in the Grand Unified Theories (GUT)~\cite{Murayama:2002jq,Meloni:2017cig,Chen:2005qm}. Because neutrinos generally take part in these theories~\cite{Panotopoulos:2021ttt}, their features could ensure experimental consequences that might be consistent or ruled out by particular theoretical models in GUT. 
In addition, self-interacting neutrinos are significant in BSM physics because they have the potential to provide important explanations for some still unknowns in neutrino physics, such as neutrino mass generation~\cite{Arguelles:2019xgp}. 

As a result, it can be concluded that the study of the Natural and CW inflation considering two different formulations of gravity: namely, metric and Palatini (especially in the non-minimal coupling to gravity scenario where the two formulations differ) are highly motivated in cosmological models with self-interacting neutrinos which substantially affect the bounds on the inflationary parameter $n_s$ (i.e., the scalar spectral index) from cosmological data.

In this work, we, for the first time, study the predictions of non-minimally coupled Natural and Coleman-Weinberg inflation potentials in metric and Palatini formulations and the predictions of non-standard neutrino self-interactions in an extended $\Lambda$CDM model in the $n_s - r$ plane, together. In this work, we consider only moderate strength interactions among $\tau$-neutrinos, $\nu_{\tau}$. This moderately interacting mode is denoted by MI$\nu$. The reason to consider only the interaction among $\nu_{\tau}$, and that too only the moderate strength interaction, is that there are strong constraints from particle physics on other possible kinds of such neutrino self-interactions with a heavy mediator, like flavor-universal neutrino self-interactions, or interaction among electron neutrinos or muon neutrinos, and even on strong interactions among $\tau$-neutrinos \cite{Blinov:2019gcj,Brdar:2020nbj,Lyu:2020lps,Berbig:2020wve}, and such interactions are essentially ruled out by particle physics experiments. However, moderate strength interactions among tau neutrinos are not ruled out by particle physics experiments and they provide similar goodness-of-fit (and Bayesian evidence) to the data as the $\Lambda$CDM model with non-interacting massive neutrinos \cite{choudhury2022massive}. Thus, there is no particular reason to favour the vanilla $\Lambda$CDM model over an extended cosmology incorporating moderately interacting tau neutrinos.  

We emphasize here that our cosmological data analysis involves assumption of a general primordial scalar power spectrum of the following form: $P_s (k) = A_s (k/k_*)^{n_s -1}$ and a general tensor power spectrum of the following form $P_t (k) = A_t (k/k_*)^{n_t}$, where $A_s$ and $A_t$ are the amplitudes of the primordial scalar and tensor fluctuations, and $n_s$ and $n_t$ are the primordial scalar and tensor spectral indices. We also impose the single-field slow-roll inflation consistency relation $n_t = - r/8$. So apart from the single-field slow-roll assumption, our cosmological model involves no other information about a particular inflationary model. Thus the bounds on $n_s$ and $r \equiv A_t/A_s$ from the cosmological data analysis are applicable to inflationary models of all types of potentials (and with or without non-minimal coupling) as long as they adhere to the single-field and slow-roll condition. This is a standard practice that is followed by the Planck collaboration as well \cite{Planck:2018jri} (see Fig 8 of this paper). It is only after generating the $n_s-r$ contour plots in the neutrino self-interaction model that we compare the contour plots with the $n_s-r$ predictions from particular inflationary models (Natural and Coleman-Weinberg inflations in this case) with various non-minimal coupling values in metric and Palatini formulations. Thus, treatment of the inflationary models are not clubbed together with cosmological analysis.

The Standard Model describes the three types of neutrinos — $\nu_e$, $\nu_{\mu}$, and $\nu_{\tau}$, and their weak interactions; nevertheless, it does not explain why these three types exist, why neutrinos have mass, and why their masses differ. A method to construct a UV complete model only with tau neutrino self-interactions should consider the concepts beyond the Standard Model, such as extensions involving new particles, forces, or dimensions. We note here that, as described in section 3, we motivate the neutrino self-interactions from the majoron model of neutrino mass generation, where the majoron appears as a Goldstone boson as the $U(1)_{B-L}$ symmetry is spontaneously broken and the majoron couples to the neutrinos via the Yukawa interaction. In general, the cosmological results presented in this paper are also valid if the mediator is a heavy gauge boson instead of a scalar (the majoron) (see \cite{Berryman:2022hds, DeGouvea:2019wpf} for detailed discussions on model-building in this field). 

However,  if one only couples a single generation of neutrinos to a gauge boson, one would generate anomalies \cite{Laha:2013xua, Dror:2017ehi,Dror:2017nsg}.  Such anomalies will lead to infinities in various processes.  The way to get around this is to postulate that such models are effective in nature, and a UV complete model will take care of the infinities at the higher scales.  One also requires that the additional particles generated by these UV complete models will have no imprint on the Cosmology. For example, suppose there is a gauge boson, $V$, coupling only to the tau neutrino.  In this case, the decay rate of $W$ boson will include the process $W \rightarrow \tau \nu_\tau V$ \cite{Laha:2013xua}.  In this case, one can show that the decay rate goes as (energy/$m_V)^2$, where $m_V$ is the mass of the gauge boson \cite{Dror:2017ehi,Dror:2017nsg}. Now, a decay rate which goes as (energy/$m_V)^2$ is anomalous since it will diverge as energy increases. Thus, this model can only be valid upto a certain energy scale.  One possible method to cancel this divergence is to introduce some new physics which will cancel the divergence; say for example there is another heavy particle which is emitted from the charged lepton which will compensate for this divergence \cite{Dror:2017ehi}.  In our scenario, we assume that this extra heavy particle does not affect cosmology and other experimental constraints. From cosmological perspective, this extra particle must be heavy and very short lived, so that various constraints do not affect it. Thus, it is quite possible to build realistic particle physics models of self-interactions among a single neutrino species. Building such a model is, however, beyond the scope of this work.  

The paper is organized as follows: We first describe the non-minimally coupled inflation in section \ref{nmc} for Metric and Palatini formulations, and inflationary parameters in section \ref{paramet}. Afterward, the considered potentials in this work, Coleman-Weinberg (CW) and Natural inflation potentials are discussed in sections \ref{cw} and \ref{nat}, respectively. In section \ref{modanaly}, we present the cosmological model and analysis methodology including perturbation equations, datasets, and parameter sampling as the subsections, and show our numerical results in section \ref{numeric}. Finally, in section \ref{result}, we discuss our results and conclude the paper.

\section{Non-minimally coupled inflation}\label{nmc}

Assuming we have a non-minimally coupled scalar field $\phi$ with a canonical kinetic term, and a potential $V_J(\phi)$, the form of Jordan frame action is described with the following form ~\cite{Jackiw:1974cv, Freedman:1974ze} 

\begin{eqnarray}\label{nonminimal_action}
S_J = \int \mathrm{d}^4x \sqrt{-g}\Big(\frac{1}{2}F(\phi) g^{\mu\nu}R_{\mu\nu}(\Gamma) - \frac{1}{2} g^{\mu\nu}\partial_{\mu}\phi\partial_{\nu}\phi -V_J(\phi)\Big),
\end{eqnarray}
where $J$ illustrates that the action is written in the Jordan frame. $F(\phi)$ describes a non-minimal coupling function, and $\phi$ indicates the inflaton. In addition, $V_J(\phi)$ is the potential term, which is given in the Jordan frame. Also, $R_{\mu\nu}$ corresponds to the Ricci tensor, which has the following form
\begin{equation}\label{Riccitensor}
R_{\mu\nu}=\partial_{\sigma}\Gamma_{\mu \nu}^{\sigma}-\partial_{\mu}\Gamma_{\sigma \nu}^{\sigma}+\Gamma_{\mu \nu}^{\rho}\Gamma_{\sigma \rho}^{\sigma}-\Gamma_{\sigma \nu}^{\rho}\Gamma^{\sigma}_{\mu \rho}.
\end{equation}
By using a metric tensor function, in the Metric formulation one can describe the connection called the Levi-Civita connection, ${\bar{\varGamma}={\bar{\varGamma}}(g^{\mu\nu})}$, with the following form
\begin{equation} \label{vargammametric}
\bar{\varGamma}_{\mu\nu}^{\lambda}=\frac{1}{2}g^{\lambda \rho} (\partial_{\mu}g_{\nu \rho}+\partial_{\nu}g_{\rho \mu}-\partial_{\rho}g_{\mu\nu}).
\end{equation}
Contrary to Metric formulation, in the Palatini formalism, the connection $\varGamma$ and $g_{\mu \nu}$ are described as independent variables, as well as with the presumption of torsion-free connection, i.e. $\varGamma_{\mu\nu}^{\lambda}=\varGamma_{\nu\mu }^{\lambda}$. If one solves the equations of motion, the following form can be acquired \cite{bauer2008inflation}
\begin{eqnarray}\label{vargammapalatini}
\Gamma^{\lambda}_{\mu\nu} = \overline{\Gamma}^{\lambda}_{\mu\nu}
+ \delta^{\lambda}_{\mu} \partial_{\nu} \omega(\phi) +
\delta^{\lambda}_{\nu} \partial_{\mu} \omega(\phi)
- g_{\mu \nu} \partial^{\lambda} \omega(\phi),
\end{eqnarray}
here, $\omega(\phi)$ has the form in terms of $F(\phi)$, and it is given by
\begin{eqnarray}
\label{omega}
\omega\left(\phi\right)=\ln\sqrt{\frac{F(\phi)}{M^2_{P}}},
\end{eqnarray}
where $M_{P}=(8 \pi G)^{-1/2}$, where $G$ is the gravitational constant. 

After an inflationary epoch, $F(\phi)\to1$ or $\phi\to0$. In this work, two different types of inflationary potentials are considered. One of them is the well-known inflation potential, which is related to the symmetry-breaking in the early universe, it is the Coleman-Weinberg inflation potential. Another one is Natural inflation which gives a plausible explication of the flatness of the inflationary potential, and it is described as the axion-like potential, thus this type of potential is very important from the particle physics viewpoint since from the spontaneously broken global symmetry, it comes out as a pseudo-Nambu-Goldstone boson. In this work, we present the inflationary predictions for both of these potentials in non-minimal coupling in the light of massive neutrino interactions.

\subsection{Inflationary parameters}\label{paramet}
The inflationary predictions can be calculated in the Einstein frame ($E$). Using Weyl rescaling, $g_E^{\mu\nu}=g^{\mu\nu}F(\phi)$, it is possible to switch from the Jordan frame to the Einstein frame. The Einstein frame action has the following form~\cite{Fujii:2003pa}
\begin{eqnarray}\label{einsteinframe}
S_E = \int \mathrm{d}^4x \sqrt{-g_{E}}\Big(\frac{1}{2}g_E^{\mu\nu}R_{E, \mu \nu}(\Gamma)-\frac{1}{2Z(\phi)}\, g_E^{\mu\nu} \partial_{\mu}\phi\partial_{\nu}\phi - V_E(\phi) \Big),
\end{eqnarray}
here, the $Z(\phi)$ term in the kinetic part of this action has different forms for the Metric and Palatini formulations. These forms can be defined separately as follows
\begin{eqnarray} \label{Zphi}
Z^{-1}(\phi)=\frac{3}{2}\frac{F'(\phi)^2}{F(\phi)^2}+\frac{1}{F(\phi)} \rightarrow Metric, \ \ \ Z^{-1}(\phi)=\frac{1}{F(\phi)} \rightarrow Palatini, 
\end{eqnarray}
where, $F'\equiv\mathrm{d}F/\mathrm{d}\phi$. In addition to this, the Einstein frame potential, $V_E(\phi)$ is described in terms of $F(\phi)$ and this has the following form 
\begin{equation} \label{Zphi2}
V_E(\phi)=\frac{V_J(\phi)}{F(\phi)^2}.
\end{equation}
One can make the field redefinition with the usage of the following expression
\begin{equation}\label{redefine}
\mathrm{d}\chi=\frac{\mathrm{d}\phi}{\sqrt{Z(\phi)}},
\end{equation}
applying this to the field redefinition, the Einstein frame action can be written in terms of the minimally coupled scalar field $\chi$ and the canonical kinetic term. By using eq. \eqref{redefine}, Einstein frame action with regard to $\chi$ can be found with the following form
\begin{eqnarray}\label{einsteinframe2}
S_E = \int \mathrm{d}^4x \sqrt{-g_{E}}\Big(\frac{1}{2}g_E^{\mu\nu}R_{E}(\Gamma)
-\frac{1}{2}\, g_E^{\mu\nu} \partial_{\mu}\chi\partial_{\nu}\chi - V_E(\chi) \Big).
\end{eqnarray}
Once the Einstein frame potential is written with the canonical scalar field $\chi$, by operating the slow-roll parameters, inflationary predictions, $n_s, r$ can be described accordingly~\cite{lyth2009primordial}. The slow-roll parameters regarding $\chi$ take the following forms
\begin{equation}\label{slowroll1}
\epsilon =\frac{M^2_{P}}{2}\left( \frac{V_{\chi} }{V}\right) ^{2}\,, \quad
\eta = M^2_{P}\frac{V_{\chi\chi} }{V}  \,,
\end{equation}
where the subscripts $\chi$'s represent the derivatives. Within the slow-roll approximation, inflationary parameters, $n_s, r$ are as follows
\begin{eqnarray}\label{nsralpha1}
n_s = 1 - 6 \epsilon + 2 \eta \,,\quad
r = 16 \epsilon,
\end{eqnarray}
here, $n_s$ is the spectral index, $r$ is the tensor-to-scalar ratio. Also, using the slow-roll approximation, the expression of the number of e-folds is in the form
\begin{equation} \label{efold1}
N_*=\frac{1}{M^2_{P}}\int^{\chi_*}_{\chi_e}\frac{V\rm{d}\chi}{V_{\chi}}\,,\end{equation}
where, the subscript ``$_*$'' indicates the quantities at the scale, which corresponds to $k_*$ that exited the horizon. In addition to this, $\chi_e$ is the inflaton value at the end of the inflationary era, one can find its value by using this equation, $\epsilon(\chi_e) = 1$. The number of e-folds takes the value, which equals approximately 60.

In terms of $\chi$, the curvature perturbation amplitude has the following form
\begin{equation} \label{perturb1}
\Delta_\mathcal{R}^2=\frac{1}{12\pi^2 M^6_{P}}\frac{V^3}{V_{\chi}^2},
\end{equation}
which should be matched with the value from the Planck outcomes~\cite{aghanim2020planck}, $\Delta_\mathcal{R}^2\approx 2.1\times10^{-9}$, considering the pivot scale, $k_* = 0.05$ Mpc$^{-1}$.

Besides that, we also present the slow-roll parameters with regard to the original field, $\phi$. For this, we need to modify the form of slow-roll parameters, which are given in terms of $\chi$ above. In our numerical calculations, we use the new forms of slow-roll parameters with $\phi$ to be able to compute easily the inflationary potentials in terms of $\phi$ for general values of free parameters, such as $\xi$, within the inflation potential forms. Otherwise, it is not simple to compute the inflationary predictions in a wide range of free parameters. By using eq. \eqref{redefine}, eq. \eqref{slowroll1} can be written in terms of $\phi$~\cite{linde2011observational} as follows 
\begin{eqnarray}\label{slowroll2}  
\epsilon=Z\epsilon_{\phi}\,,\quad
\eta=Z\eta_{\phi}+{\rm sgn}(V')Z'\sqrt{\frac{\epsilon_{\phi}}{2}},
\end{eqnarray}
where we describe
\begin{equation}
\epsilon_{\phi} =\frac{1}{2}\left( \frac{V^{\prime} }{V}\right) ^{2}\,, \quad
\eta_{\phi} = \frac{V^{\prime \prime} }{V}.
\end{equation}
Moreover, equations \eqref{efold1} and \eqref{perturb1} are written as to $\phi$ in the following forms
\begin{eqnarray}\label{perturb2}
N_*&=&\rm{sgn}(V')\int^{\phi_*}_{\phi_e}\frac{\mathrm{d}\phi}{Z(\phi)\sqrt{2\epsilon_{\phi}}}\,,\\
\label{deltaR} \Delta_\mathcal{R}&=&\frac{1}{2\sqrt{3}\pi}\frac{V^{3/2}}{\sqrt{Z}|V^{\prime}|}\,.
\end{eqnarray}

Throughout this work, we suppose the standard thermal history after the inflationary era. Concerning this, the inflationary predictions for the considered potentials will be calculated. With this consideration, $N_*$ takes the form~\cite{liddle2003long, Maji:2022jzu} as follows
\begin{eqnarray} \label{efolds}
N_*\approx61.5+\frac12\ln\frac{\rho_*}{M^4_{P}}-\frac{1}{3(1+\omega_r)}\ln\frac{\rho_e}{M^4_{P}} 
+\Big(\frac{1}{3(1+\omega_r)}-\frac14\Big)\ln\frac{\rho_r}{M^4_{P}}\,,
\end{eqnarray}
for the pivot scale $k_* = 0.05$ Mpc$^{-1}$. In addition, within the form of $N_*$, $\rho_{e}=(3/2)V(\phi_{e})$ indicates the energy density at the end of inflation. $\rho_r=(\pi^2/30)g_*T^4_r$ represents the energy density at the end of reheating, here $T_r$ indicates the reheating temperature, as well as $\rho_{*} \approx V(\phi_*)$ is the energy density at which the scales coincide with $k_*$, which exited the horizon, and $\rho_*$ can be defined by applying eq. \eqref{deltaR}, then it has the following form
\begin{equation}
    \rho_{*} = \frac{3 \pi^2\Delta^2_\mathcal{R} r }{2}.
\end{equation}
Furthermore, $\omega_r$ corresponds to the equation-of-state parameter during reheating. In this work, we use $\omega_r=1/3$, which defines the assumption of instant reheating. With the selection of $\omega_r=1/3$, we eliminate the dependence of the reheating temperature in the $N_*$ definition,  which is given in eq. \eqref{efolds}. In our numerical analysis, for the number of e-folds, we use eq. \eqref{efolds} with $\omega_r=1/3$, as well as we use the units in the reduced Planck scale $M_P=1/\sqrt{8\pi G}\approx2.43\times10^{18}\text{ GeV}$ and it will be taken as equal to 1.

\subsection{Coleman-Weinberg inflation}
\label{cw}
Since new inflation models were proposed in the early eighties, the Coleman-Weinberg mechanism has been related to symmetry-breaking in the very early universe ~\cite{linde1982new,albrecht1982cosmology, coleman1973radiative, albrecht1985realization, linde1990particle, Shafi:1983bd, Lazarides:1984pq}. In the Jordan frame, the form of effective Coleman-Weinberg potential is as follows 
\begin{equation}\label{cwpot}
V_J(\phi) = A \phi^4\left[\ln \left(\frac{\phi}{v}\right)-\frac{1}{4}\right]+\frac{A v^4}{4},
\end{equation}
where $v$ indicates the vacuum expectation value (VEV) of the inflaton. This form of potential can be described in the Einstein frame by using the non-minimal coupling function, $F(\phi)$. 

In this work, for the Coleman-Weinberg potential, we use the form of $F(\phi)$ by following the ref.~\cite{Bostan:2018evz} as follows
\begin{eqnarray}\label{nonminimal_coupls1}
F(\phi)=1+\xi(\phi^2-v^2).
\end{eqnarray}
Thus, we take into account the Coleman-Weinberg potential for two different cases:
\begin{itemize}
    \item Above the VEV: $\phi>v$,
    \item Below the VEV: $\phi<v$.
\end{itemize}

Using eq. \eqref{nonminimal_coupls1}, the form of Coleman-Weinberg potential in the Einstein frame can be obtained as follows

\begin{equation}\label{cwpotE}
V_E(\phi) = \frac{A \phi^4\left[\ln \left(\frac{\phi}{v}\right)-\frac{1}{4}\right]+\frac{A v^4}{4}}{\Big[1+\xi(\phi^2-v^2)\Big]^2}.
\end{equation}

This potential with minimal coupling, $\xi=0$, is already taken into account in the refs.~\cite{shafi2006coleman, barenboim2014coleman, kannike2014embedding, shafi2015primordial,Smith:2008pf,Rehman:2008qs,Okada:2014lxa}. The potential form in equation \eqref{cwpotE} for both Metric and Palatini formalism was considered previously, for instance ref.~\cite{Bostan:2018evz} for Metric, ref.~\cite{Bostan:2020pnb} for Palatini with details. Also, this form of Einstein frame Coleman-Weinberg potential with non-minimal coupling to gravity for $w_{r}=0$ and different reheating temperature values was considered with detail in~\cite{Maji:2022jzu}, they use Metric formulation of gravity. On the other hand, in this work, we present our results for the non-minimally coupled Coleman-Weinberg potential, which is defined in eq. \eqref{cwpotE}, in both Metric and Palatini formulations for $w_{r}=1/3$ in the light of massive neutrino interactions. It is important to note that here, ref.~\cite{choudhury2022massive} examined the Coleman-Weinberg potential in the light of massive neutrino interactions in a minimal coupling case, so by taking $\xi=0$.
\subsection{Natural inflation}
\label{nat}  
Natural inflation was introduced for the first time to find the key to \textit{fine-tuning} of inflation \cite{martin2014encyclopaedia}. Besides that, Natural inflation gives a plausible elucidation to the inflaton potentials' flatness, which is essential for the smooth form of inflationary potential. Also, this potential is very crucial because it can be described by the axion-like inflaton~\cite{freese1990natural, reyimuaji2021natural,adams1993natural}, and it makes this potential such an attractive archetype in particle physics because of the spontaneously broken global symmetry, the inflaton field, $\phi$ naturally appears as a pseudo-Nambu-Goldstone boson~\cite{freese1990natural}. In Natural inflation models, $\phi$ is the axion-like inflaton, as well as the model, has a cosine-type periodic potential. 

The potential form of Natural inflation in the Jordan frame is as follows
\begin{equation}\label{nipot}
V_J(\phi)=V_0\left[1+ \cos \left(\frac{\phi}{f}\right)\right],
\end{equation}
where $f$ is the symmetry-breaking scale. In literature, many studies have examined the Natural Inflation potential for minimal ($\xi=0$) and non-minimal couplings ($\xi \neq 0$), such as \cite{Bostan:2022swq, Reyimuaji:2020goi, Zhou:2022ovp, Stein:2021uge, German:2021jer, Simeon:2020lkd, AlHallak:2022gbv}. In this work, we investigate the non-minimally coupled Natural inflation potential by using a non-minimal coupling function, which is in the form
\begin{eqnarray}\label{nonminimal_coupls2}
F(\phi)=1+\xi \phi^2.
\end{eqnarray}
Thus, the non-minimally coupled Natural inflation in the Einstein frame can be written as follows
\begin{equation}\label{nipotE}
V_E(\phi)=\frac{V_0\left[1+ \cos \left(\frac{\phi}{f}\right)\right]}{\left(1+\xi \phi^2\right)^2}.
\end{equation}
In this work, we demonstrate our results for the Natural inflation, which is defined in the Einstein frame in eq. \eqref{nipotE}, for both Metric and Palatini formulations by taking $w_{r}=1/3$ in the light of massive neutrino interactions. In ref.  \cite{choudhury2022massive}, Natural inflation was also studied in the light of massive neutrino interactions but in a minimal coupling case ($\xi=0$).

\section{Cosmological model and analysis methodology} \label{modanaly}

Neutrinos are massless in the Standard Model of particle physics, but terrestrial neutrino oscillation experiments have confirmed that at least two out of the three active neutrino mass eigenstates are non-zero. There are a plethora of models for neutrino mass generation. For this particular work, we incorporate the majoron model where the neutrinos are Majorana particles, and the $U(1)_{B-L}$ \cite{Gelmini:1980re,Choi:1991aa,Acker:1992eh,Chikashige:1980ui,Georgi:1981pg} symmetry is spontaneously broken. This leads to a new Goldstone boson, the majoron, denoted by $\Phi$. It couples to the neutrinos via the Yukawa interaction~\cite{Oldengott:2017fhy,Oldengott:2014qra},

\begin{equation}\label{interlag}
\mathcal{L}_{\rm int} = g_{ij} \bar{\nu_i} \nu_j \Phi + h_{ij}  \bar{\nu_i} \gamma_5 \nu_j \Phi.
\end{equation} 
Here $\nu_i$ is a left-handed neutrino Majorana spinor, and $g_{ij}$ and $h_{ij}$ are the scalar and pseudo-scalar coupling matrices, respectively. The indices $i,j$ are labels for the neutrino mass eigenstates. In general, interactions of this kind are not limited to only the majoron-like model of neutrino mass generation. For instance, $\phi$ can be linked to the dark sector~\cite{Barenboim:2019tux}.

In this paper, we consider a flavor-specific interaction scenario (only 1 neutrino species interacting), specifically the $\tau-$neutrino. This is because other scenarios, like the flavor-universal interaction scenario or flavor-specific interactions among $\nu_e$ and $\nu_{\mu}$ are strongly constrained by particle physics experiments, and self-interactions among only the $\nu_{\tau}$ are allowed, that too only the moderately interacting mode (denoted by MI$\nu$) \cite{Blinov:2019gcj,Brdar:2020nbj,Lyu:2020lps,Berbig:2020wve}. Here we have a diagonal $g_{ij} = g\delta_{kk}\delta_{ij}$, where $k$ is either 1, 2, or 3 (no sum over $k$ is implied) in the mass basis, where only one diagonal term is non-zero. Note that a diagonal $g_{ij}$ in the flavor basis with only one non-zero component $g_{\tau\tau}$ does not imply a diagonal $g_{ij}$ in the mass basis with only one non-zero component. Nonetheless, the non-diagonal terms or other diagonal terms in the mass-basis $g_{ij}$ are expected to be small given that the bound on the sum of neutrino masses from cosmology is quite strong even in the presence of neutrino self-interactions \cite{RoyChoudhury:2020dmd}, i.e., we are essentially dealing with quite small neutrino masses. Thus we expect this approximation to be reasonable. Also, since we are only interested in a still-viable particle physics model, we only include the $\logg$ parameter range for the MI$\nu$ mode in our analysis. We denote this model as ``1$\nu-$interacting, MI$\nu$."
 
In this work, the mass of the scalar, $m_{\Phi}$, is considered to be much larger than the energies of neutrinos during the CMB epoch. This allows us to reasonably regard the interaction to be an effective 4-fermion interaction for the CMB epoch and later. A mass of  $m_{\Phi} > 1$ keV is enough to ensure this \cite{Blinov:2019gcj}, and the $\Phi$ particles would have decayed away when the temperature falls below 1 keV. However, to avoid constraints from Big Bang Nucleosynthesis, one might consider $m_{\Phi} > 1$ MeV. We note here that such a scenario is not restricted to a scalar particle. In fact, all the results and conclusions in this paper will be applicable for a heavy vector-boson as well, since neutrino self-interactions with a heavy vector-boson can also be cast as a 4-fermion interaction when the temperature drops below the mass of the vector-boson \cite{Ohlsson:2012kf,Archidiacono:2013dua}. 

Now the interaction Lagrangian in equation \eqref{interlag} can be written as a $\nu \nu \rightarrow \nu \nu$ self-interaction. The self-interaction rate per particle $\Gamma \sim g^4 T_{\nu}^5/m_{\phi}^4 = G_{\rm eff}^2 T_{\nu}^5$, where $G_{\rm eff} = g^2/m_{\phi}^2$ is the effective self-coupling~\cite{Oldengott:2017fhy}. In this given scenario, the neutrinos decouple from the primordial plasma, as usual, at a temperature of $T \sim 1$ MeV, when the weak interaction rate falls below the Hubble rate, i.e. $\Gamma_{\rm W} < H$, with $\Gamma_{\rm W} \sim G_{\rm W}^2 T_{\nu}^5$. Here $G_{\rm W} \simeq 1.166 \times 10^{-11} \textrm{MeV}^{-2}$ is the standard Fermi constant. But even after this decoupling, the neutrinos continue to scatter among themselves if $G_{\rm eff} > G_{\rm W}$. This self-scattering continues until the self-interaction rate $\Gamma$ falls below the Hubble expansion rate $H$, and only after this, the neutrinos will free-stream, unlike the standard case where they start free-streaming right after decoupling from the primordial plasma. Thus, by increasing  $G_{\rm eff}$, one can further delay the neutrino free-streaming. Very strong interactions like $G_{\rm eff} \simeq 10^9 G_{\rm W}$ can impede free-streaming till matter radiation equality. 

\subsection{Cosmological model and perturbation equations}

The cosmological model of interest here is an extended $\Lambda$CDM model that includes the tensor-to-scalar ratio $r_{0.05}$, and sum of neutrino masses $\sum m_{\nu}$, effective number of relativistic species $N_{\rm eff}$, and the logarithm of the interaction strength $\textrm{log}_{10} \left[G_{\rm eff} \textrm{MeV}^2\right]$.

Since only one of the neutrinos is interacting, we denote it as the 1$\nu$-interacting model. This cosmological model can be represented by the same following parameter vector: 

\begin{equation}\label{eq:model}
\theta = \{\Omega_{\rm c} h^2,\Omega_{\rm b} h^2,100\theta_{MC},\tau,
\ln(10^{10}A_{s}),n_{s}, r_{0.05}, \sum m_\nu, N_{\rm eff}, \textrm{log}_{10} \left[G_{\rm eff} \textrm{MeV}^2\right]\}.
\end{equation}

Here, the first six parameters are associated with the standard $\Lambda$CDM model. $\Omega_{\rm c} h^2$ and $\Omega_{\rm b} h^2$ are the physical densities at present ($z=0$) for cold dark matter (CDM) and baryons respectively, $100\theta_{MC}$ is a parameter used by CosmoMC ~\cite{Lewis:2002ah,Lewis:2013hha} as an approximation for the angular size of the sound horizon, $\theta_s$. We have $\tau$ as the optical depth of reionization and $\ln(10^{10}A_s)$ and $n_s$ are the amplitude and spectral index of the primordial scalar fluctuations, respectively, at a pivot scale of $k_*=0.05 \rm ~Mpc^{-1}$. 
Apart from $A_s$ and $n_s$, the tensor-to-scalar ratio $r$ is another important parameter for inflationary models. We also use a pivot scale of $k_*=0.05 \rm ~Mpc^{-1}$ for $r$, and hence we denote it as $r_{0.05}$. 
 
In our analyses, we use the degenerate hierarchy of neutrino masses, i.e., all the neutrino masses are equal. One can write: $m_\nu = \frac{1}{3}\sum m_\nu$, where $m_\nu$ is the mass of each neutrino. Presently, the 95\% bound on $\sum m_\nu$ is close to 0.1 eV \cite{RoyChoudhury:2018gay,RoyChoudhury:2018vnm,RoyChoudhury:2019hls,Vagnozzi:2017ovm,Vagnozzi:2018jhn,Giusarma:2018jei,Tanseri:2022zfe}, but there is no conclusive evidence for a preference for the normal or inverted hierarchy of the masses of neutrinos \cite{RoyChoudhury:2019hls,Gariazzo:2022ahe,Gariazzo:2018pei,Heavens:2018adv,Lattanzi:2017ubx,Gerbino:2016ehw}, and thus the degenerate hierarchy is okay to be used. This is actually true even for near-future cosmological datasets \cite{Archidiacono:2020dvx,Mahony:2019fyb,MoradinezhadDizgah:2021upg}. Also, we use a flat prior on $\textrm{log}_{10} \left[G_{\rm eff} \textrm{MeV}^2\right]$ instead of $G_{\rm eff}$ because it enables us to vary the parameter over multiple orders of magnitude. 

We emphasize here that while doing the numerical analyses, we distribute the $N_{\rm eff}$ equally among the 3 neutrinos. So in the 1$\nu$-interacting model, only $N_{\rm eff}/3$ corresponds to the self-interacting neutrino species, and the rest is associated with free-streaming neutrinos.

The priors on the model parameters are listed in table~\ref{tab:priors}. Since we are only interested in the Moderately Interacting Mode (MI$\nu$) we use the following prior range: $-5.5\rightarrow -2.3$.   

We also perform our analysis in the $\Lambda\textrm{CDM} +r_{0.05}$ model, as we want to analyze the predictions of the inflationary models against both the $\Lambda\textrm{CDM} +r_{0.05}$ model and the ``1$\nu-$interacting, MI$\nu$" model.

\begin{table}[t]
\caption{Uniform priors for all the cosmological model parameters.}
\label{tab:priors}
\begin{center}
\begin{tabular}{lr@{$\,\to\,$}l}
\hline
 Parameter & \multicolumn{2}{c}{Prior}\\
\hline
$\Omega_{\rm b}h^2$ & $0.019$ & $0.025$\\
$\Omega_{\rm c}h^2$ & $0.095$ & $0.145$\\
$100\theta_{MC}$ & $1.03$ & $1.05$\\
$\tau$ & $0.01$ & $0.1$\\
$n_s$ & $0.885$ & $1.04$\\
$\ln{(10^{10} A_s)}$ & $2.5$ & $3.7$\\
$r_{0.05}$ & $0$ & $0.3$ \\
$\sum m_\nu$ [eV] &  $0.005$ & $1$\\
$N_{\rm eff}$ & $2$ & $5$\\
$\textrm{log}_{10} \left[G_{\rm eff} \textrm{MeV}^2\right]$  & $-5.5$ & $-2.3$\\
\hline
\end{tabular}
\end{center}
\end{table}

We modify the cosmological perturbation equations of neutrinos in the CAMB code \cite{Lewis:1999bs}. The modifications to the perturbation equations apply only to one of the three species. The background equations remain the same as the non-interacting case because the neutrinos are only self-interacting, i.e., there is no energy transfer between the neutrino sector and any other sector. Also, the heavy mediator decays away much before photon decoupling. 

To include the self-interaction in the neutrino perturbation equations in CAMB, we use the relaxation time approximation (RTA) that was first introduced in this context in~\cite{Hannestad:2000gt} (and first used for the treatment of self-interactions in light neutrinos in~\cite{Hannestad:2004qu}). RTA was shown to be very accurate in comparison to the exact collisional Boltzmann equations, in \cite{Oldengott:2017fhy}. Note that we have incorporated the modifications to both scalar and tensor perturbation equations.

In the scalar perturbation equations of neutrinos, these self-interactions cause damping in the Boltzmann hierarchy for multipoles $\ell\geq 2$ due to the scattering. In the synchronous gauge, the collisional Boltzmann hierarchy for massive neutrino scalar perturbations is given by (following the notation in \cite{Ma:1995ey}),
\begin{eqnarray}
\label{eq:boltzman}
\dot{\Psi}_0 &=& -{qk\over \epsilon}\Psi_1
+{1\over 6}\dot{h} {d\ln f_0\over d\ln q}
\,, \nonumber\\
\dot{\Psi}_1 &=& {qk\over 3\epsilon} \left(\Psi_0
- 2 \Psi_2 \right) \,, \nonumber\\
\dot{\Psi}_2 &=& {qk\over 5\epsilon} \left(
2\Psi_1 - 3\Psi_3 \right)
- \left( {1\over15}\dot{h} + {2\over5} \dot{\eta} \right)
{d\ln f_0\over d\ln q} + \alpha_2 \dot{\tau}_\nu \Psi_2\,,\\
\dot{\Psi}_l &=& {qk \over (2l+1)\epsilon} \left[ l\Psi_{l-1}
- (l+1)\Psi_{l+1} \right] + \alpha_\ell \dot{\tau}_\nu \Psi_l \,,
\quad l \geq 3 \,, \nonumber
\end{eqnarray}

where $\alpha_\ell \dot{\tau}_\nu \Psi_l$ are the damping terms for $l\geq 2$. Here, $\dot \tau_\nu \equiv -a G_{\rm eff}^2 T_\nu^5$ is the opacity for the neutrino self-interactions with a heavy mediator, and $\alpha_l$ ($l>1$) are coefficients of order unity that depend on the interaction model. We use $\alpha_l$ values from equation\ 2.9 in~\cite{Oldengott:2017fhy} for the scalar mediator, i.e., we use $\alpha_2 = 0.40$, $\alpha_3 = 0.43$, $\alpha_4 = 0.46$,  $\alpha_5 = 0.47$, $\alpha_{l \geq 6} = 0.48$. For neutrino tensor perturbation equations we go through a similar procedure and incorporate similar damping terms to the perturbation equations in the CAMB code \cite{Lewis:1999bs}. However, we use $\alpha_l = 1$ ($l>1$), instead of including model-specific values, since these model-dependent coefficients for tensor perturbation equations require separate elaborate calculations. We have verified that when we vary $\alpha_l$ from 0.4 to 1, the CMB B-mode spectrum sourced by the primordial tensor perturbations goes through only a small change. Thus, fixing all $\alpha_l = 1$ in the neutrino tensor perturbation equations is only going to produce very minor shifts in the value of $\logg$ and thus, is not of any major concern.

We also incorporate a tight coupling approximation (TCA) in our code, in the very early universe. In TCA, only the two lowest moments are non-zero. We use TCA because the collisional Boltzmann equations for neutrinos are difficult to solve in the very early universe. This approximation is switched off quite early (when $|\dot \tau_\nu|/\mathcal{H} < 1000$, where $\mathcal{H}$ is the conformal Hubble parameter) so that it does not bias our results.

\subsection{Datasets}
We make use of the full CMB temperature and polarisation data (i.e. TT, TE, EE + lowE) from the Planck 2018 public data release~\cite{Planck:2018vyg}. Here, TT signifies the low-$l$ and high-$l$ temperature power spectra, TE signifies the high-$l$ temperature and E-mode polarisation cross-spectra, EE signifies the high-$l$ E-mode polarisation spectra, and lowE the low-$l$ E mode polarisation spectra. Additionally, we also include the B-mode CMB power spectra data from the BICEP2/Keck array public data release~\cite{BICEP:2021xfz} that includes observations up to 2018. In addition to the CMB data, we include Planck 2018 CMB lensing~\cite{Planck:2018lbu}, BAO and RSD measurements from SDSS-III BOSS DR12~\cite{BOSS:2016wmc}, additional BAO measurements from MGS \cite{Ross:2014qpa} and 6dFGS \cite{Beutler:2011hx}, and SNe~Ia luminosity distance measurements from the Pantheon sample~\cite{Pan-STARRS1:2017jku}. 

\subsection{Parameter sampling}
We use the public nested sampling package Polychord~\cite{Handley:2015vkr,Handley:2015fda} added to CosmoMC~\cite{Lewis:2002ah,Lewis:2013hha}, known as CosmoChord \cite{Handley}. We use 2000 live points and boost\_posterior = 0 to properly sample the parameter space. We use HMcode \cite{Mead:2020vgs} (included with the CosmoChord package) to handle the non-linear part of the cosmological evolution. We use GetDist~\cite{Lewis:2019xzd} to generate the bounds on the parameters and the posterior plots.

\subsection{Effect of neutrino self-interactions on the CMB BB spectrum}

\begin{figure}[h]
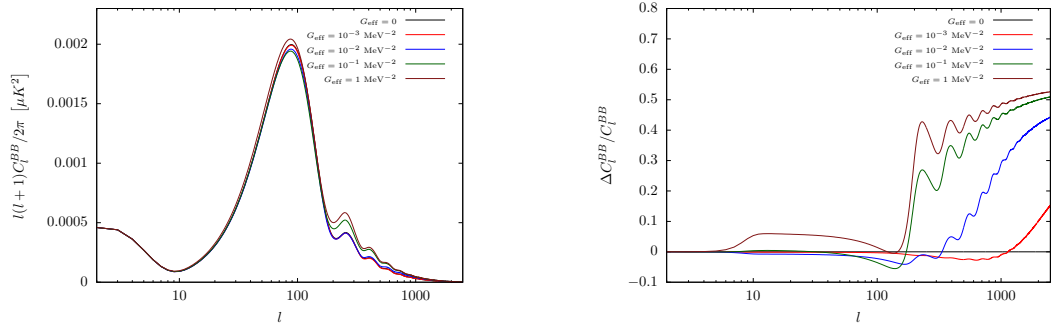

        \resizebox{.5\linewidth}{!}{\parbox{\linewidth}{\input{3nu_int_pp.tex}}}\resizebox{.5\linewidth}{!}{\parbox{\linewidth}{\input{3nu_int_pp_compare.tex}}}	
\caption{\label{fig:0} The left panel shows the primordial CMB B-mode power spectrum $C_l^{BB}$ (appropriately scaled for better visualisation) for different values of $G_{\rm eff}$ for all 3 neutrinos interacting, where other cosmological parameters remain fixed to suitable values (particularly, tensor-to-scalar ratio is fixed to $r_{0.05} = 0.03$). The right panel shows the ratio $\Delta C_l^{BB}/C_l^{BB}$, where the denominator is the B-mode power spectrum for $G_{\rm eff} = 0$, and  $\Delta C_l^{BB} \equiv C_l^{'BB} - C_l^{BB}$, where $C_l^{'BB}$ are the B-mode power spectra for values of $G_{\rm eff}$ as specified by the legends in the top right corner. } 
\end{figure}

The effect of neutrino self-interactions (similar to the heavy mediator case studied here) on the CMB B-mode (primordial) spectrum has been previously studied in \cite{Loverde:2022wih}. The effect of DM-neutrino interactions on the primordial B-mode of CMB are expected to be similar to the neutrino self-interactions \cite{Ghosh:2017jdy}. In figure \ref{fig:0}, we show the effect of neutrino self-interactions for various different values of the coupling strength $G_{\rm eff}$ for the case where all 3 neutrinos are interacting. Similar to the previous studies, we find that strong neutrino self-interactions can enhance the CMB B-mode of the power-spectrum by as much as 50\% in the $l>100$ regime. The effect of only one neutrino species interacting will be proportionately smaller. To detect these effects in the CMB B-mode, there will be a need of not only precise measurement of the CMB B-mode power spectrum, but also very advanced delensing techniques \cite{Seljak:2003pn}, since the $l>100$ region is expected to be dominated by CMB lensing B-modes which are sourced  at low redshifts from lensing of primordial E-modes by the matter structure. \\

As far as the sensitivity of the current datasets is concerned (especially BICEP/Keck \cite{BICEP:2021xfz}), we find that introduction of neutrino self-interactions has negligible effects on the bounds on $r_{0.05}$. Thus, as far as inflationary parameters are concerned, the main effect of neutrino self-interactions is through the effect on the scalar spectral index $n_s$, as we had also seen previously in \cite{choudhury2022massive}.

\section{Numerical results} \label{numeric}
In this section, we discuss the non-minimally coupled Natural and Coleman-Weinberg potentials in the light of massive neutrino interactions and show our numerical results. In our numerical analysis, we use the potential forms which are given in equations. \eqref{cwpotE} and \eqref{nipotE} for the non-minimally coupled Coleman-Weinberg and Natural inflation potentials, respectively. In addition, in our analysis, we use the e-fold number that is presented in eq. \eqref{efolds}, it provides us with an assumption of the standard thermal history after the end of inflation. It is good to mention that we set $\omega_r=1/3$, which corresponds to the instant reheating, with this assumption we ignore the dependence of reheating temperature in our analysis. In our numerical calculations for predictions from inflationary models and also in our analysis of cosmological data, we take a pivot scale $k_* = 0.05$ Mpc$^{-1}$. Also, we present our results considering an MI$\nu$ model in the 1$\nu$-interacting scenario, this indicates that the self-interaction is constrained to solely one flavor of neutrinos (specifically, $\tau$ neutrinos). The details about the models and datasets are given in section \ref{modanaly}.

\begin{figure}[tbp]
    \includegraphics[width=0.47\textwidth]{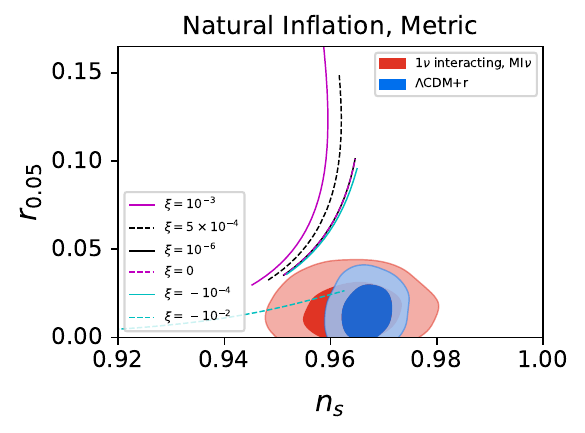}
    \hfill
    \includegraphics[width=0.47\textwidth]{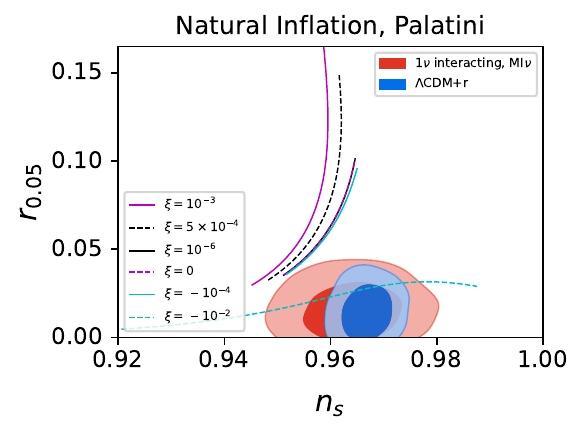}
    \caption{\label{fig:1} The predictions of the inflationary parameters in the $n_s-r$ plane for the Natural inflation potential in the Metric (left panel) and Palatini (right panel) formulations for various values of $\xi$ and the 68\% and 95\% C.L. contour plots for the 1$\nu$-interacting MI$\nu$ model (shaded red) and the $\Lambda$CDM$+r$ model (shaded blue). }
\end{figure}

First, we begin discussing our results for the inflationary predictions of the non-minimally coupled Natural inflation potential in Metric and Palatini formulations, the related outcomes are shown in the figure \ref{fig:1}. According to our results, $n_s-r$ predictions for the selected $\xi$ values in this study, except $\xi=-10^{-2}$, of both formulations, are ruled out at 2$\sigma$ for the  MI$\nu$ model in the 1$\nu$-interacting scenario. It can be concluded that aside from $\xi=-10^{-2}$, the inflationary predictions of the non-minimally coupled Natural inflation potential for both formulations cannot enter into either 1$\sigma$ or 2$\sigma$ confidence regions in 1$\nu$-interacting  MI$\nu$ model for the $\xi$ values we choose in this study. Also, $n_s-r$ predictions for $\xi \ll 1$, such as $\xi=10^{-6}$ overlap with the results of $\xi=0$ (minimal coupling case), thus we can say that the $n_s-r$ predictions of $\xi \ll 1$ are not able to accommodate at confidence regions for our neutrino interaction scenario. On the other hand, for $\xi=-10^{-2}$, the inflationary predictions can be inside the 1$\sigma$ region for the $f$ values, $f\sim3.5$ for Metric, $f\sim3.05$ for Palatini in the 1$\nu$-interacting MI$\nu$ model. Except $\xi=-10^{-2}$, for each selected $\xi$ value in this study, we can emphasize that the inflationary predictions of two formulations are very close to each other for the non-minimally coupled Natural inflation potential, and none of these results are in the confidence regions for the 1$\nu$-interacting MI$\nu$ model. For $\xi=-10^{-2}$, on the other hand, for the large $f$ values, the inflationary predictions have different patterns for the Metric and Palatini formulations, for instance, the inflationary predictions stay in 1$\sigma$ at $f\sim3.5$ for Metric formulation, while for the Palatini, the results are outside the confidence regions for $f\sim3.2$ with the increase of $n_s$, reaching $n_s \sim 0.987$.

\begin{figure}[tbp]
  \centering
    \includegraphics[width=0.47\textwidth]{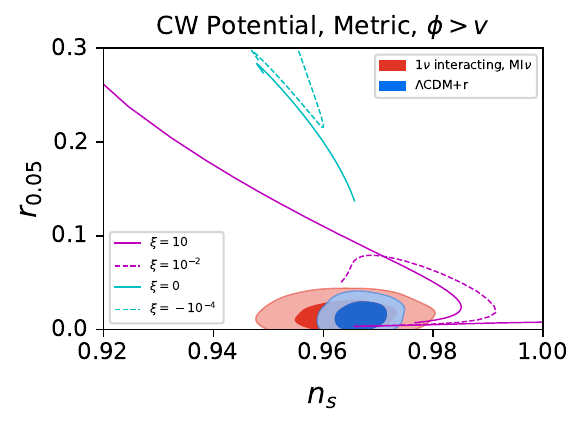}
    \caption{\label{fig:2} The predictions of the inflationary parameters in the $n_s-r$ plane for the Coleman-Weinberg inflation potential in the Metric formulation for various values of $\xi$ in the $\phi > v$ case, and the 68\% and 95\% C.L. contour plots for the 1$\nu$-interacting MI$\nu$ model (shaded red) and the $\Lambda$CDM$+r$ model (shaded blue).}
\end{figure}

\begin{figure}[tbp]
\includegraphics[width=0.47\textwidth]{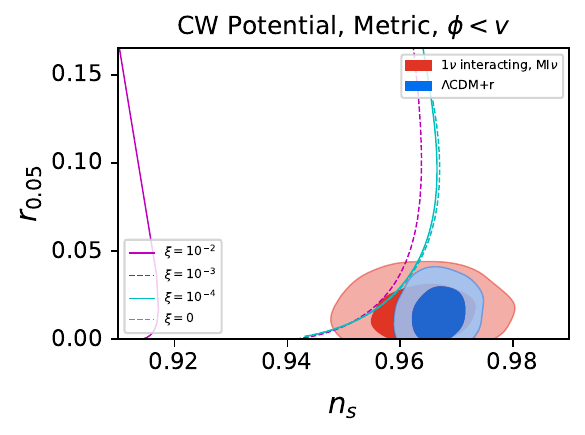}\hfill\includegraphics[width=0.47\textwidth]{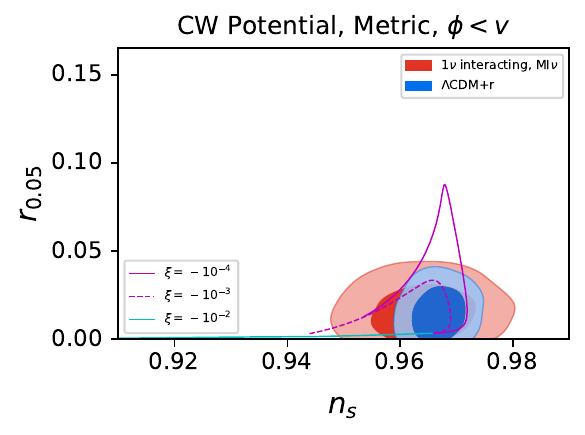}
    \caption{\label{fig:3} The predictions of the inflationary parameters in the $n_s-r$ plane for the Coleman-Weinberg inflation potential in the Metric formulation for various values of $\xi$ in the $\phi < v$ case, and the 68\% and 95\% C.L. contour plots for the 1$\nu$-interacting MI$\nu$ model (shaded red) and the $\Lambda$CDM$+r$ model (shaded blue). }
\end{figure}

Secondly, we present our results for the non-minimally coupled Coleman-Weinberg inflation potential in Metric formulation. Figures \ref{fig:2} and \ref{fig:3} present the results of $\phi>v$ and $\phi<v$ cases, respectively. In the Metric formulation, for $\phi>v$ case, the inflationary predictions for $\xi=10^{-2}$ and $\xi=10$ can be inside the confidence regions for the MI$\nu$ model in the 1$\nu$-interacting scenario but for $\xi=0$ and $\xi=-10^{-4}$, the predictions are ruled out according to this scenario. Furthermore, for $\phi<v$ case, the predictions of $\xi=10^{-2}$ are also ruled out in this scenario, on the other hand, for $\xi=10^{-3}$, $\xi=10^{-4}$ and $\xi=0$ values, the $n_s-r$ can enter into 1$\sigma$ confidence region. Similarly, for $\xi<0$ values, we show the results of $\xi=-10^{-4}$, $\xi=-10^{-3}$ and $\xi=-10^{-2}$, the inflationary predictions can be also inside the 1$\sigma$ of this scenario. 

\begin{figure}[tbp]
  \centering
\includegraphics[width=0.47\textwidth]{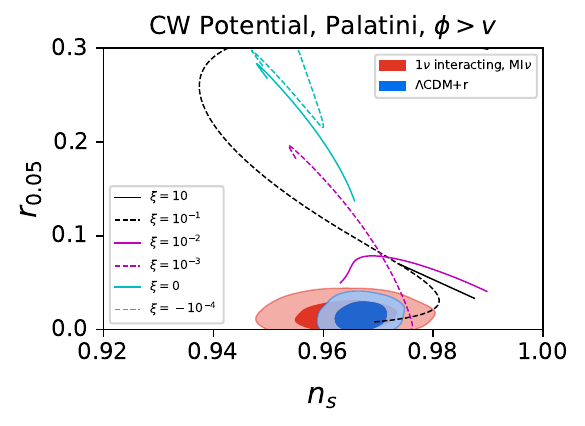}
    \caption{\label{fig:4} The predictions of the inflationary parameters in the $n_s-r$ plane for the Coleman-Weinberg inflation potential in the Palatini formulation for various values of $\xi$ in the $\phi > v$ case, and the 68\% and 95\% C.L. contour plots for the 1$\nu$-interacting MI$\nu$ model (shaded red) and the $\Lambda$CDM$+r$ model (shaded blue).}
\end{figure}

\begin{figure}[tbp]  

\includegraphics[width=0.47\textwidth]{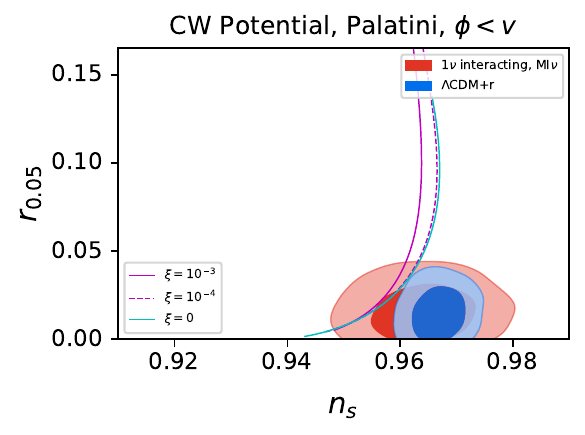}\hfill\includegraphics[width=0.47\textwidth]{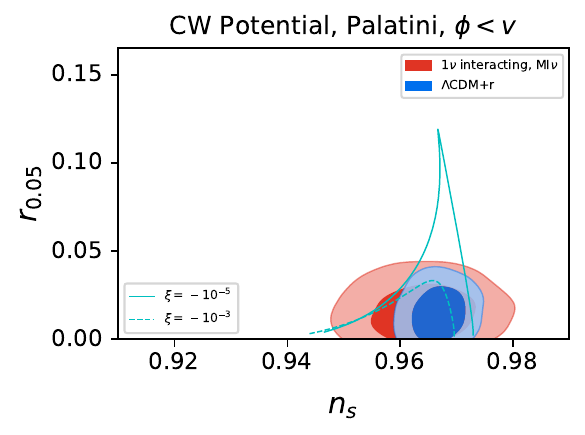}
    \caption{\label{fig:5} The predictions of the inflationary parameters in the $n_s-r$ plane for the Coleman-Weinberg inflation potential in the Palatini formulation for various values of $\xi$ in the $\phi < v$ case, and the 68\% and 95\% C.L. contour plots for the 1$\nu$-interacting MI$\nu$ model (shaded red) and the $\Lambda$CDM$+r$ model (shaded blue). }
\end{figure}

Lastly, we show the results for the non-minimally coupled Coleman-Weinberg potential in the Palatini formulation.  Figures \ref{fig:4} and \ref{fig:5} present the results of $\phi>v$ and $\phi<v$ cases, respectively. In the Palatini formulation, for the $\phi>v$ case, the inflationary predictions for $\xi=10^{-1}$ can be accommodated in the 1$\sigma$ for 1$\nu$-interacting MI$\nu$ scenario but for $\xi=10^{-3}$, the results are only within the 2$\sigma$ confidence region for this scenario. However, for $\xi=10$, $\xi=10^{-2}$, $\xi=-10^{-4}$ and $\xi=0$, the predictions cannot enter into even the 2$\sigma$ confidence regions of this scenario at all. For $\phi<v$ values, the predictions for all selected $\xi$ values in this work can be inside the 1$\sigma$ confidence region for the 1$\nu$-interacting MI$\nu$ model. 
\begin{itemize}
\item For $\phi>v$ case:
 $\rightarrow$ For the Coleman-Weinberg potential, we show that the inflationary predictions saturate the linear potential limit for $\xi\gtrsim 10^{-1}$ values in the Palatini formulation, this result is consistent with the studies ~\cite{Bostan:2020pnb, Racioppi:2017spw}. This limit can be accommodated in the 1$\sigma$ confidence region for the 1$\nu$-interacting case in the MI$\nu$ model. 

\item For $\phi<v$ case:
  $\rightarrow$ The pattern of the inflationary predictions for the non-minimally coupled Coleman-Weinberg inflation potential is almost the same for both Metric and Palatini formulations for each selected $\lvert \xi \rvert$ value in this work. Thus, the situations for entering the confidence interval for the 1$\nu$-interacting MI$\nu$ model are almost similar for these $\xi$ values. Also, the results in Metric formulation show that $n_s$ values have large shift ($n_s\sim 0.92$) for $\xi=10^{-2}$, thus the $n_s-r$ predictions of $\xi=10^{-2}$ are outside the confidence regions in the 1$\nu$-interacting MI$\nu$ model.

\end{itemize}

\section{Conclusions} \label{result}

In this work, we have analyzed the inflationary predictions of the Coleman-Weinberg and Natural inflation potentials with non-minimal coupling and the inflationary parameter predictions of the cosmological model 1$\nu$-interacting MI$\nu$, i.e., an extended $\Lambda$CDM model with one neutrino species having a moderate strength self-interaction. We have presented the predictions of these potentials for both the Metric and Palatini formulations of gravity. It is important to remind here that for the symmetry-breaking related Coleman-Weinberg potential, we have considered that the inflaton takes a non-zero $v$ after the inflationary era, therefore we have shown our results of this potential for two different cases: $\phi>v$ and $\phi<v$. 

After reviewing our theoretical background, as well as the cosmological model, analysis methodology, and datasets in this work, we presented our numerical results in the paper. For both of the considered inflationary potentials, by taking neutrino interactions into account, we have shown whether the inflationary predictions ($n_s$ and $r$) are compatible or not with the recent cosmological data. We studied the predictions of the inflationary potentials and the predictions of both the $\Lambda\textrm{CDM} +r_{0.05}$ model and the ``1$\nu-$interacting, MI$\nu$" model, which we previously explained.

We have found that the predictions of non-minimally coupled Natural inflation potential can be inside the confidence regions only for $\xi=-10^{-2}$ of both Metric and Palatini formulations in our 1$\nu-$interacting, MI$\nu$ model (and also in the minimal $\Lambda$CDM$+r_{0.05}$ model). Moreover, we have shown that the inflationary predictions for $\xi \ll 1$ overlap with the results of the minimal coupling case, and these results are ruled out at more than 2$\sigma$ for our neutrino interaction scenario. The inflationary predictions of $\xi=-10^{-2}$ are inside the 1$\sigma$ region at $f\sim3.5$ ($\sim3.05$) for Metric (Palatini) formulations in the 1$\nu$-interacting MI$\nu$ model. Also, for $\xi=-10^{-2}$, for the large $f$ values, the inflationary predictions have different forms for the Metric and Palatini formulations, for example, the inflationary predictions remain in 1$\sigma$ at $f\sim3.5$ for the Metric formulation, while for the Palatini, the results cannot be inside the confidence regions for $f\sim3.2$ with the increment of $n_s$, approaching $n_s \sim 0.987$. It is worth mentioning here that excluding $\xi=-10^{-2}$, the inflationary parameters of two formulations for chosen $\xi$ values in this study are very close to each other for the Natural inflation potential, therefore the inflationary predictions ($n_s-r$) for these $\xi$ values are almost similar, and these predictions are outside of the 2$\sigma$ confidence limits for the 1$\nu-$interacting, MI$\nu$ model.

Furthermore, for the non-minimally coupled Coleman-Weinberg potential in Metric formulation, the inflationary parameters for $\xi=10^{-2}$ and $\xi=10$ can be within the 2$\sigma$ region in the 1$\nu$-interacting MI$\nu$ scenario, but for the predictions of $\xi=0$ and $\xi=-10^{-4}$, the predictions are ruled out at 2$\sigma$ according to the 1$\nu$-interacting MI$\nu$ model. In addition, for the $\phi<v$ case, the predictions of $\xi=10^{-2}$ are ruled out in this scenario. On the other hand, for $\xi=10^{-3}$, $\xi=10^{-4}$ and $\xi=0$, the $n_s-r$ can be inside the 1$\sigma$ region. Similarly, for $\xi<0$ values that we have displayed, i.e., the inflationary predictions of $\xi=-10^{-4}$, $\xi=-10^{-3}$ and $\xi=-10^{-2}$ can also be within the 1$\sigma$ confidence region for the 1$\nu-$interacting, MI$\nu$ model. 

Lastly, for the non-minimally coupled Coleman-Weinberg inflation potential in the Palatini formulation, we have shown that the inflationary predictions for $\xi=10^{-1}$ can remain within the 1$\sigma$ region in the 1$\nu$-interacting scenario. It is good to mention here that for the Palatini Coleman-Weinberg inflation potential, the inflationary predictions saturate the linear potential limit for $\xi\gtrsim 10^{-1}$ values, at which the parameters stay within 1$\sigma$ for 1$\nu$-interacting MI$\nu$ model. On the other hand, the inflationary parameters for $\xi=10^{-3}$ are inside the 2$\sigma$ of 1$\nu-$interacting, MI$\nu$ model but for $\xi=10$, $\xi=10^{-2}$, $\xi=-10^{-4}$ and $\xi=0$, the predictions do not enter into the 2$\sigma$ confidence region for our neutrino interaction scenario, at all. For the $\phi<v$ case, the predictions for all the selected $\xi$ values in this study, $\xi=10^{-3}$, $\xi=10^{-4}$, $\xi=-10^{-3}$, $\xi=-10^{-4}$ and the minimal coupling case ($\xi=0$), can be inside the 1$\sigma$ confidence region for the 1$\nu$-interacting  MI$\nu$ scenario. 
     
While we use cosmological data to constrain the inflationary parameters in a cosmological model that has self-interaction among neutrinos, a detection of neutrino self-interactions is yet to be confirmed by particle physics experiments. However, if such an interaction is detected in the future by experiments like DUNE \cite{DUNE:2020ypp, Bakhti:2018avv, Berryman:2019dme} and IceCube \cite{IceCube:2024kel, IceCube-Gen2:2020qha}, then it would imply that inflationary models like Natural inflation and Coleman-Weinberg inflation (which are considered ruled out by the data in the minimal $\Lambda$CDM model) will remain afloat as viable inflationary theories. Additionally, certain values of non-minimal coupling studied in this paper, which are ruled out by the data in the minimal $\Lambda$CDM model, will also remain viable in the event of detection of such self-interactions among neutrinos.

\acknowledgments
We thank Ranjan Laha and Sudip Jana for helpful discussions. We thank the referee for useful suggestions. We acknowledge the use of the HPC facility at Aarhus University (http://www.cscaa.dk/) where a part of the numerical analyses was done.

\bibliographystyle{jhep}
\bibliography{biblio}




\end{document}